\documentstyle[12pt]{article}
\begin{document}
\begin{center}
{\Large \bf
{Brane-Universe in Six Dimensions with Two Times}
}\\
\vskip 0.5cm
{ Merab GOGBERASHVILI } \\
\vskip 0.3cm
{\it
 {Institute of Physics, Georgian Academy of Sciences}\\
 {6 Tamarashvili Str., Tbilisi 380077, Georgia}\\
 {(E-mail: gogber@hotmail.com)}} \\
\vskip 0.5cm

{\Large \bf Abstract}\\
\vskip 0.5cm
\quotation
{\small Brane-Universe model embedded in 6-dimensional space-time with
the signature (2+4) is considered. A matter is gravitationally trapped
in three space dimensions, but both time-like directions are open.
Choosing of the dimension and the signature of the model is initiated
with the conformal symmetry for massless particles and any point in our
world can be (1+1) string-like object.}
\endquotation
\end{center}
\vskip 0.5cm

In conventional Kaluza-Klein's picture extra dimensions are curled up to
unobservable size. Last years models with large extra dimensions become
popular (see for example \cite{ADD,RS}). Those approaches also do not
contradict to present time experiments \cite{OW}. Main obstacles in
progress of such models were how to confine a matter inside
brane-Universe and to explain observed 4-dimensional Newton's law. In
the papers \cite{G1,G2,G3,G4} we had introduced possible mechanism of
overcoming these problems. Special solution of multi-dimensional
Einstein's equations \cite{G1,G4,V}, responsible for gravitational
trapping of a matter, provides effective 4-dimensional Newton's law on
the brane. In this model it can be explained also non-observation of
cosmological constant in four dimensions \cite{G1,G4,RuS}.

In our previous papers \cite{G1,G2,G3,G4} for simplicity 5-dimensional
model with the signature (1+4) was considered. In general the question
about the number of dimensions and signature is open. We consider that a
matter is trapped on the brane by gravitation. So it is natural to
assume that for massless case (it means weakest coupling with gravity)
symmetries of sub-manifold can be restored. It is well known that in the
zero-mass limit main equations of physics are invariant under fifteen
parameter nonlinear conformal transformations. From the other hand a
long time ago it was discovered that conformal group can be written as a
linear Lorentz type transformations in 6-dimensional space-time with the
signature (2+4) (for these subjects see for example \cite{PR}). Another
indication that real number of space-like dimensions of Universe can be
four, could be $O(4)$ symmetry of the solution of Shrodinger equations
for the hydrogen atom.

Theories with extra time-like dimensions have been a subject of interest
for some time (the latest article about large extra time-like dimensions
is \cite{DGS}, see also \cite{BDM,Vo}). Kaluza-Klein 6-dimensional model
with two times, but with compact extra dimensions, was investigated
before in \cite{P,I}. Necessity of two times follow also from string
theory (F-theory) \cite{Va}.
There exist another two groups of 6-dimensional schemes, but with three
times. The schemes of the first group \cite{MR,Pa,Z} suffer from
internal inconsistencies, second \cite{C}, more sophisticated scheme is
internally consistent.

In all these papers multi time dimensions are introduced, however, not
much is known about the classical vacuum solutions of two- or
more-timing theories. It is known that theories with compact internal
time-like dimensions have several pathological features. The most
conspicuous may be the fact that excitations of the internal dimensions
have negative norm. Experimental lower bounds on possible violation of
unitarity put a limit on the maximum radius of the internal time-like
directions \cite{Y}.

After unification of space and time coordinates (but not dimensions) in
Lorents transformations many physicists intuitively are accustomed to
consider space and time dimensions identically. However, there is
principal difference among them. Main difference is that one can easily
change position, or stop in the space but everybody follows to time flow
which was began from the Big Bang. Decreasing of time flow, or shift of
time vector for any system, means backward moving in the time and then
disappearance for other observers. Another problem is that we know
several space directions and it is easy to add next ones, but we know
only one time. In case of two or more time directions the question
arises what we are measuring with our clocks. Thus ordinary methods of
trapping using in our previous papers \cite{G1,G2,G3,G4} can be not
applicable for time-like dimensions.

Possible indication of existence of extra time directions can be
non-conservation of the energy in four dimensions. This can be happening
only after interaction with the matter with another direction of time
(or energy) vector. As we mentioned above we only passively follow to
time flow, direction of which coincides with cosmological arrow of time
(possible source of different asymmetries in our world, see for example
\cite{BG1}). So it could work mechanism similar to that considered in
papers \cite{C} - after some time from the Big Bang all particles with
another directions of time had been disappeared from our view. In this
case may be no special trapping mechanism in extra times is necessary.

All points of the space in Universe are equivalent. However, we have
global zero for time coordinate, it is the moment of the Big Bang. What
we measure for any particle is the difference of two energies - energy
of the vacuum and the particle itself. When particle follows to our time
flow we notice only this difference, but to shift the time vector we
need total value of the energy corresponding to the age of Universe.
Thus if the particles with another energy vector had been disappeared
the matter of our world is trapped in our time by Plank scale.

In this paper we consider 6-dimensional Kaluza-Klein model with one
space-like and one time-like extended extra dimension. It is
generalization of 5-dimensional scheme of papers \cite{G1,G2,G3,G4} with
one brane where all matter is localized. Trapping mechanism considered
in these articles works for the case of one extra space-like dimension
and is applicable for 6-dimensional model with two times. Time-like
dimensions in our scheme are open, similar to the approach of \cite{C}.

We are looking for solution of 6-dimensional Einstein's equations with
the cosmological term $\Lambda$ in two time- and four space-like
dimensions
\begin{equation}  \label{1}
^6R_{AB}-\frac 12g_{AB}~^6R = -\Lambda g_{AB} + G\tau _{AB}~~.
\end{equation}
Here $^6R_{AB}$ and $G$ are 6-dimensional Ricci tensor and gravitational
constant and big Latin indices $ A, B,... = 0, 1, 2, 3, 5, 6 $.

Energy-momentum tensor for brane-Universe with the signature (2+3)
embedded in 6-dimensional space-time with signature (2+4) is taken in
the form
\begin{equation} \label{2}
\tau_{\mu\nu} = g_{\mu\nu}\sigma \delta (\frac{x^6}{\epsilon}), ~~
\tau_{55} = \sigma \delta (\frac{x^6}{\epsilon}),~~ \tau_{66} = 0~~.
\end{equation}
Here $\sigma$ is brane tension, $\delta (x^6/\epsilon )$ - delta
function and $\epsilon \sim \sqrt{\Lambda}$ - brane width in extra
space-like dimension $x^6$. Greek indices $\alpha, \beta ... = 0, 1, 2,
3$ numerate coordinates in four dimensions. This form of the brane
energy-momentum follows from kink-like solution of scalar field
equations without introducing of coupling with gravity. In general it is
difficult to separate the energy of the brane and gravitation field and
we must consider more complicate model.

In (\ref{1}) we choose negative sign for cosmological constant $\Lambda
$. Canceling mechanism of papers \cite{G1,G2,G3,G4} for space-like extra
dimension works only for the negative $\Lambda $. Solution with positive
$\Lambda $ corresponds to trapping in extra time and is considered at
the end of the paper.

In many Kaluza-Klein models with large extra dimensions localization of
particular quantum fields on the brane is investigated (latest paper in
this direction is \cite{BG}). In our model \cite{G1,G2,G3,G4} matter is
trapped on the brane by gravitation. The source of anti-gravity
responsible for this trapping can be negative multi-dimensional
cosmological constant. Since gravity is universal field we don't need
any other classical source for localization of a matter on the brane.
Exact mechanism for different fields is difficult to find because of
problems with consideration of quantum fields in curve space-time (see
for example \cite{BD}). Even in four dimensions only few exactly
solvable models exist. Our approach is more general. Gravitation field
is not localized on the brane, however, by canceling mechanism we have
ordinary Newton's law. Gravitational potential has minimum on the brane
and all particles having coupling with gravity are sitting there.

To keep the width of brane-Universe during expansion, it means for the
stabile localization of the mater on the brane, as in paper \cite{G3} we
look for the solution of (\ref{1}) with zero extra momentum
\begin{equation}  \label{3}
P_i=\int T_i^A dS_A = 0~~~.
\end{equation}
Small Latin indices $i, j, k, ... = 5, 6$ numerate coordinates of extra
dimensions. Here $ T_A^B = t_A^B + \tau _A^B $ is total energy momentum
tensor of matter fields $\tau _A^B $ and gravitational field itself
\begin{equation} \label{4}
t_{A}^{B} = \frac{1}{2G}[g^{BD}\partial_{A}\Gamma^{E}_{DE} -
g^{ED}\partial_{A}\Gamma^{B}_{DE} + \delta^{B}_{A}(~^6R - 2\Lambda)]~~.
\end{equation}
Type of matter fields is not important now. Some particular cases are
considered in the paper \cite{G3}.

In this article we don't want to touch the old problem with the energy
of gravitation and as in the paper \cite{G3} choose $T_A^B$ in the form
of so called Lorentz energy-momentum complex
\begin{equation}  \label{5}
T_A^B=\frac 1{2G\sqrt{g}}\partial _CX_A^{BC}~~,
\end{equation}
where
\begin{equation}  \label{6}
X_A^{BC}=-X_A^{CB}=\sqrt{g}[g^{BD}g^{CE}(\partial _Dg_{AE}-\partial
_Eg_{AD})]~~.
\end{equation}
This form is convenient, since in this case energy-momentum tensor of
gravitational field $t_{A}^{B}$ coincides with canonical energy-momentum
tensor (\ref{4}) received from Hilbert's form of the gravitational
Lagrangian $ L_g = \sqrt{g}(~^6R - 2\Lambda)$.

To satisfy the stability condition (\ref{3}), components of the
energy-momentum tensor on the solutions must satisfy the condition
\begin{equation}  \label{7}
T_i^A =0~~.
\end{equation}
From $(i\alpha )$ component of this relation and (\ref{4}) we find
$\partial _{i}\Gamma^{\alpha}_{\mu\nu} = 0 $. Thus simple solution of
(\ref{7}) is
\begin{equation}  \label{8}
g_{i\alpha }=0~~,~~~g_{\alpha\beta}= \lambda (x^i) \eta_{\alpha\beta}
(x^{\nu})~~,
\end{equation}
where $\eta_{\alpha\beta}(x^{\nu})$ is ordinary 4-dimensional metric
tensor and $\lambda (x^i)$ is arbitrary function of extra coordinates.
Solution (\ref{8}) which we received from stability conditions, is
similar with the anzats of \cite{RuS}.

Stability condition (\ref{7}) for the case of diagonal metric tensor of
extra dimensions $(g_{56} = 0)$ has the form
\begin{equation}  \label{9}
\partial_5(g^{55}g^{66}\partial_5g_{66})
=\partial_6(g^{55}g^{66}\partial_6g_{55}) = 0~~.
\end{equation}
One of the solutions of this system is
\begin{equation}  \label{10}
g_{55}= exp(cx^6)~~,~~~g_{66} = - 1~~,
\end{equation}
where $c$ is integration constant.

Using (\ref{8}) one can find decomposition of Einstein's equations (see
also \cite{RS})
\begin{eqnarray}  \label{11}
R_{\alpha\beta} - \frac{1}{2}g_{\alpha\beta}(D_iD^i\lambda +
\frac{1}{2\lambda}D_i\lambda D^i\lambda ) = \frac{1}{2}
g_{\alpha\beta}[\Lambda - \frac{1}{2}G \sigma \delta
(\frac{x^6}{\epsilon})]~~, \nonumber \\
R_{ij} - \frac{2}{\lambda}(D_iD_j\lambda - \frac{1}{2\lambda}D_i\lambda
D_j\lambda) = \frac{1}{2} g_{ij}[\Lambda - \frac{5}{2}G \sigma \delta
(\frac{x^6}{\epsilon})]~~ .
\end{eqnarray}
Here $R_{ij}$ and $D_i$ are correspondingly Ricci tensor and the
covariant derivative in extra space-time with the metric tensor $
g_{ij}$. Ricci tensor in four dimensions $R_{\alpha\beta}$ is
constructed from $ g_{\alpha\beta} = \lambda (x^i) \eta_{\alpha\beta}
(x^{\nu})$ in a standard way.

Using (\ref{10}) and the properties of the step function $H(x^5)$
\begin{equation}  \label{12}
|x|^{\prime} = H(x) - H(-x)~~,~~~ H(x)^{\prime} = \delta(x) =
\frac{1}{|\epsilon |}\delta (\frac{x}{\epsilon)}
\end{equation}
where prime denotes derivative, one can show that system (\ref{11}) has
the trapping solution
\begin{equation}  \label{13}
\lambda = g_{55} = exp(c|x^6|)~~,~~~g_{66} = -1~~.
\end{equation}
The integration constant here has the value
\begin{equation}  \label{14}
c =\sqrt{\frac{2\Lambda}{5}} = - \frac{G\sigma \epsilon}{4}~~.
\end{equation}
This formula also contains necessary relation between the brane tension
$\sigma $ and 6-dimensional cosmological constant $\Lambda $.

For this solution in four dimensions we have ordinary Einstein's
equations without the cosmological term
\begin{equation}  \label{15}
R_{\alpha\beta} - \frac{1}{2}\eta_{\alpha\beta}R = 0 ~~,
\end{equation}
which is function of only 4-dimensional metric tensor
$\eta_{\alpha\beta}(x^{\nu})$. After adding of 4-dimensional source at
the right hand of (\ref{15}) one can show that as in 5-dimensional case
\cite{G1,G4} a matter is trapped in three space with ordinary Newton's
low, while now we have unobservable extra open time direction.

From (\ref{14}) we see that constant $c$ is real only for our choice of
sign of $\Lambda $. Also we noticed that for branes with positive
tension constant $c$ is negative and function $\lambda $ decreasing fare
from the brane, as in papers \cite{RS}. For the case of negative $\sigma
$ exponential factor in (\ref{13}) is positive as in papers
\cite{G1,G4,V} and gravitational potential has minimum on the brane.

In this paper and our previous articles \cite{G1,G2,G3,G4}, in contrast
with the approach of \cite{RS}, we consider only one brane. Interactions
of branes with the negative and positive tensions are often considered
in Kaluza-Klein theories with the large extra dimensions. One must be
careful in this case. It is known for a long time \cite{B} that system
of negative and positive masses began to accelerate till the speed of
the light. Acceleration can destroy the branes. Even one brane with
positive tension is strange objects, since it is gravitationally
repulsive \cite{VS}. This can cause change of the time direction on the
brane \cite{BG2}, while $t^2$ is still positive. Negative tension can
change signature and thus interchanges time and space coordinates.

System (\ref{11}) with positive $\Lambda $  has similar to (\ref{13})
solution
\begin{equation}  \label{16}
\lambda = - g_{66} = exp(c|x^5|)~~,~~~g_{55} = 1~~,
\end{equation}
which corresponds to trapping in extra time, while all four space-like
coordinates are open.

At the end of the paper we want to note that in this paper 6-dimensional
model with string-like extra dimension was considered. Any point-like
particle in our world can have tail in (1+1) dimensions and we have
interesting possibility for nontrivial application of string theory.


\begin{thebibliography}{30}

\bibitem{ADD} N. Arkani-Hamed, S. Dimopoulos and G. Dvali,
              Phys. Lett. {\bf B429}, 263 (1998); Phys. Rev. {\bf D59},
086004 (1999).

\bibitem{RS} L. Rundall and R. Sundrum,
             Phys. Rev. Lett. {\bf 83}, 3370 (1999); {\bf 83}, 4690
(1999).

\bibitem{OW}  J. M.  Overduin and P. S. Wesson,
              Phys. Rept. {\bf 283}, 303 (1997).

\bibitem{G1}   M. Gogberashvili,
               hep-ph/9812296.

\bibitem{G2}   M. Gogberashvili,
               hep-ph/9812365; Europhysics Lett. {\bf 49}, (2000).

\bibitem{G3}   M. Gogberashvili,
              hep-ph/9904383; Mod. Phys. Lett. {\bf A14}, 2024 (1999).

\bibitem{G4}   M. Gogberashvili,
              hep-ph/9908347.

\bibitem{V} M. Visser,
            Phys. Lett. {\bf B159}, 22 (1985).

\bibitem{RuS} V. A. Rubakov and M. E. Shaposhnikov,
              Phys. Lett. {\bf B125}, 139 (1983).

\bibitem{PR} R. Penrose and W. Rindler,
             {\it Spinors and Space-time} (University Press of
Cambridge, Cambridge, 1986).


\bibitem{DGS} G. Dvali, G. Gabadadze and G. Senjanovic,
              hep-ph/9910207; {\it A contribution to the Yu. A. Golfand
memorial volume}, Ed. M. A. Shifman (World Scientific, 1999).

\bibitem{BDM} I. Bars, C. Deliduman and D. Minic,
             hep-th/9906223.

\bibitem{Vo} S. Vongehr,
             hep-th/9907034.

\bibitem{P}  M. Pavsic,
             Nuovo Cimento {\bf B41}, 397 (1977).

\bibitem{I}  R. L. Ingraham,
             Nouvo Cimento {\bf B46}, 1 (1978) 1; {\bf B46}, 16 (1978);
{\bf B46}, 217 (1978); {\bf B46}, 261 (1978); {\bf B47}, 157 (1978);
{\bf B50}, 233 (1979); {\bf B68}, 203 (1982); {\bf B68}, 218 (1982).

\bibitem{Va} C. Vafa,
             Nucl. Phys. {\bf B469}, 403 (1996).

\bibitem{MR}  R. Magnani and E. Recami,
              Lett. Nouvo Cimento {\bf 16}, 449 (1977).

\bibitem{Pa}  P. T. Pappas,
             Lett. Nouvo Cimento {\bf 22}, 601 (1978); Nuovo Cimento
{\bf B68}, 11 (1982).

\bibitem{Z}  G. Ziino,
             Lett. Nouvo Cimento {\bf 24}, 191 (1979).

\bibitem{C}  E. A. B. Cole,
             Nouvo Cimento {\bf B40}, 171 (1977); J. Phys. {\bf A13},
109 (1980).

\bibitem{Y}  F. J. Yndurain,
             Phys. Lett. {\bf B256}, 15 (1991).

\bibitem{BG} B. Bajc and G. Gabadadze,
              hep-th/9912232.

\bibitem{BD} N. D. Birrell and P. C. W. Davies,
            {\it Quantum Fields in Curved Space} (University Press of
Cambridge, Cambridge, 1982).

\bibitem{BG1} A. Barnaveli and M. Gogberashvili,
              Phys. Lett. {\bf B316}, 57 (1993).

\bibitem{B} H. Bondi,
            Rev. Mod. Phys. {\bf 29}, 423 (1957).

\bibitem{VS} A. Vilenkin and E. P. S. Shellard,
            {\it Cosmic Strings and Other Topological Defects}
(University Press of Cambridge, Cambridge, 1994).

\bibitem{BG2} A. Barnaveli and M. Gogberashvili,
             hep-ph/9505412; Gen. Rel. Grav. {\bf 26}, 1117 (1994); {\it
New Frontiers in Gravitation} (Hadron Press, Palm Harbor, 1996).

\end{thebibliography}
\end{document}